\documentclass[aps,prl,superscriptaddress,twocolumn,floats]{revtex4}
\usepackage{graphicx}
\usepackage{amsmath}
\usepackage{amssymb}
\usepackage{bm}
\usepackage{amsmath}
\usepackage{mathtools}
\usepackage[x11names,svgnames]{xcolor}

\newcommand\s{\sigma}

\newcommand\Gv{\mathbf{G}}
\newcommand\Sv{\mathbf{S}}
\newcommand\Kv{\mathbf{K}}
\newcommand\Tv{\mathbf{T}}
\newcommand\kv{\mathbf{k}}
\newcommand\rv{\mathbf{r}}

\newcommand\kvt{\mathbf{\tilde k}}
\newcommand\er{\mathrm{e}}
\newcommand\etal{\textit{et al.}}
\renewcommand\L{\langle}
\newcommand\R{\rangle}

\newcommand\dr{\mathrm{d}}

\renewcommand\L{\langle}

\renewcommand\vec[1]{\mathbf{#1}}

\begin{document}
\title{Stable Algebraic Spin Liquid in a Hubbard Model}

\author{S. R. Hassan}
\affiliation{The Institute of Mathematical Sciences, C.I.T. Campus, Chennai 600 113, India}

\author{P.V. Sriluckshmy}
\affiliation{The Institute of Mathematical Sciences, C.I.T. Campus, Chennai 600 113, India}

\author{Sandeep K. Goyal}
\affiliation{The Institute of Mathematical Sciences, C.I.T. Campus, Chennai 600 113, India}
\affiliation{School of Chemistry and Physics, University of KwaZulu-Natal, Durban, South Africa}

\author{R. Shankar}
\affiliation{The Institute of Mathematical Sciences, C.I.T. Campus, Chennai 600 113, India}

\author{David S\'en\'echal}
\affiliation{D\'epartment de Physique and RQMP, Universit\'e de Sherbrooke, Sherbrooke, Qu\'ebec, J1K 2R1, Canada}

\date{\today}
\begin{abstract}
We show the existence of a stable algebraic spin liquid (ASL) phase in a
Hubbard model defined on a honeycomb lattice with spin-dependent hopping that
breaks time-reversal symmetry. The effective spin model is the Kitaev model for
large on-site repulsion. The gaplessness of the emergent Majorana fermions is
protected by the time-reversal invariance of this model. We prove that
the effective spin model is time-reversal invariant in the entire Mott phase, thus ensuring the stability of the ASL. The model can be physically realized in cold atom systems, and we propose experimental signals of the ASL.
\end{abstract}
\maketitle


The concept of a spin liquid as a Mott phase without any local magnetic order
was put forward by Anderson  \cite{fazekas}. Its relevance to the physics of
high-temperature superconductors\cite{Anderson:1987, baskaran1988} led to the
development of a gauge theory of spin liquids \cite{senthil2000,baskaran1988},
analogous to quantum electrodynamics(QED). The spinons are the counterpart of
electrons in QED, and the visons, another emergent excitation, are the
counterpart of the photon. Attempts at understanding the emergence of fermionic
quasiparticles in spin systems in analogy with the anyonic quasiparticles in
fractional quantum Hall systems have led to a general theory of
quantum or topological order in spin liquids\cite{wen2002}. Experimental evidence for a spin liquid ground state has been seen, for instance, in the organic
material $\kappa$-(BEDT-TTF)$_2$Cu$_2$(CN)$_3$\cite{Shimizu:2003uq}.

Algebraic spin liquids (ASL) are a special class of spin liquids with gapless Dirac-like spinons and spin correlations that decay as a power law.
Frustration in magnetic interactions and quantum fluctuations tend to prevent magnetic ordering. Thus, ASLs have primarily been studied in frustrated spin-1/2 Heisenberg antiferromagnets\cite{affleck1988, rantner2001, senthil2005} and have not been realized in an interacting fermion model until now. The ASL shows power-law decay not only for spin correlations but for many other local order parameters as well. Hence, it is intrinsically susceptible to one of them
ordering and inducing a spinon gap. Thus, any realization of this phase must be accompanied by a mechanism for ensuring its stability.

Kitaev\cite{kitaev} constructed an exactly solvable anisotropic spin-1/2 model on a honeycomb lattice that exhibits the important properties of an ASL.  It can be expressed as a model of two gapless Majorana-Dirac fermions (spinons) interacting with $Z_2$ gauge fields (visons). A remarkable feature of the model is that the magnetic flux associated with every plaquette is conserved, and as a result the visons are static. Consequently, while multispin operators that conserve flux have algebraic correlations, those which do not, including the single spin operators, are extremely short ranged \cite{saptarshi1}.  Tikhonov {\it et al.}\cite{Tikhonov:2011kx} showed that when a single spin operator is added to the Hamitonian , the spin-spin correlations become algebraic as well. The class of perturbations that can induce algebraic spin-spin correlations was classified by Mandal {\it et al.} \cite{Mandal:2011ys}, who showed that Ising and Heisenberg perturbations, which had been studied earlier \cite{Mandal:2011ys,Chaloupka:2010uq,schaffer2012}, do not induce power-law correlations.

ASLs are thus realized in a class of perturbed Kitaev models. The stability of the ASL in the Kitaev model is due to time-reversal (TR) symmetry: the two Majorana-Dirac fermions combine to form a {\em single} Dirac fermion, with an energy spectrum that cannot have a gap without breaking TR symmetry. Thus, to ensure a stable ASL phase, the perturbations must preserve this symmetry. The single spin perturbation considered by Tikhonov {\it et al.} breaks TR symmetry, and hence that model is not protected against developing a spinon gap at higher orders in perturbation theory.  An exactly solvable spin-3/2 model with algebraic spin correlations has also been constructed \cite{kivelson}.

The ASL has not yet been realized in a model of interacting fermions, though there have been speculations about the possible means of doing so\cite{hermele2007,yingran,sondhi}. It has been argued\cite{Meng:2010kx} that a short-ranged spin liquid emerges between the semi-metal and the N\'eel phases in a Hubbard model defined on a honeycomb lattice. Recent works claim otherwise\cite{sorella, jeanpaul}.  

\begin{figure}
\includegraphics[width=4.5cm]{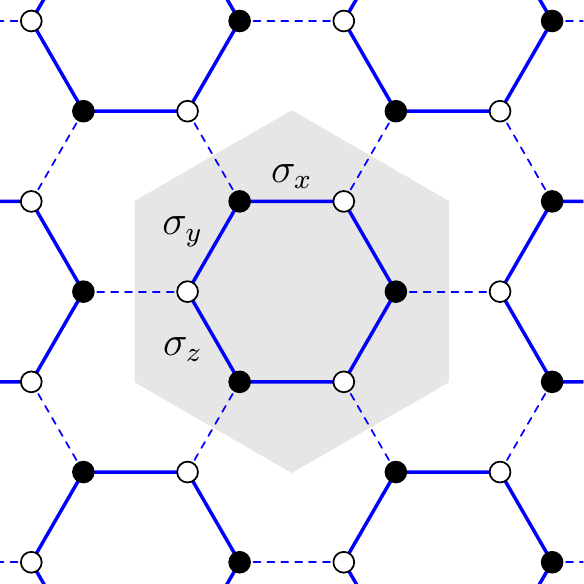}
\caption{(color online)
The honeycomb lattice with the two sublattices marked by white and black dots. The six-site cluster used in this work is shown as the shaded area. The $\sigma_i$ label the different spin-dependent hopping directions (blue solid lines), whereas the inter-cluster bonds are shown as dashed lines.
}
\label{fig:cluster}
\end{figure}

The above discussion suggests that a Hubbard model that would be described effectively by the Kitaev honeycomb spin model in the large $U$ limit is a good candidate for realizing an ASL. Such a model was proposed by Duan \etal\ as a way of realizing the Kitaev model\cite{kitaev} in cold atom systems\cite{Duan:2003dq}. This model, which we henceforth call the Kitaev-Hubbard model, has anisotropic spin-dependent hopping, which leads to the high degree of frustration in the effective spin model.  The Hamiltonian is
\begin{equation}
\label{eq:ham} 
H=-\sum_{\L ij\R_\alpha}
\left\{c^\dagger_i\left(\frac{t+t'\s_\alpha}{2}\right)
c_j+\mathrm{H.c.}\right\}+U\sum_i~n_{i\uparrow}n_{i\downarrow}, 
\end{equation}
where $c_{i\s}$ annihilates a fermion of spin projection
$\s=\uparrow,\downarrow$ at site $i$ (the spin index is implicit in the first
term), $\s_\alpha$ ($\alpha =x,y,z$) are the Pauli matrices, $n_\s\equiv
c^\dagger_\s c_\s$ is the number of fermions of spin $\s$ at site $i$, and $\L
ij\R_\alpha$ denotes the nearest-neighbor pairs in the three hopping directions
of the lattice (see Fig.~\ref{fig:cluster}). 

In the remaining part of the Letter, we analyze this model and show that there exists a phase with a charge gap and no magnetic order. We compute the next to leading order (in $t/U$) terms of the effective spin model in the Mott phase and show that they induce algebraic spin-spin correlations. The effective model has an emergent time-reversal symmetry, which we analytically prove remains intact to all orders in $t/U$. Finally, we do a mean-field calculation to show that time-reversal symmetry is not spontaneously broken in the Mott phase. Thus, we demonstrate that the model supports a {\em stable} ASL phase.

At $t'=0$, the model reduces to the simple spin- and TR-invariant, nearest-neighbor Hubbard model\cite{Meng:2010kx,sorella,jeanpaul}. The term proportional to $t'$ is a spin-dependent hopping term and breaks TR symmetry, $SU(2)$ spin symmetry, and the threefold spatial rotation symmetry of the $t'=0$ model.  It is, however, invariant under a spatial rotation of $2\pi/3$ combined with a spin rotation of $2\pi/3$ about the $(111)$ spin axis.  At $t'=t$, the one-body part of the Hamiltonian is a combination of the projection operators $\frac12(1+\s_\alpha)$.  Thus, only those electrons that are spin-polarized in the $\alpha{\rm th}$ direction can hop along the $\alpha$ bonds.  At this value of $t'$, the effective low-energy spin model, at half-filling and large $U$, is the Kitaev honeycomb model\cite{Duan:2003dq,Zhang:2007cr}. 

At $U=0$ and $t' = 1$, the single-particle spectrum of this model shows four distinct bands, each of which has a nonzero Chern number $\nu$\cite{Hassan:2012fk}. The top and bottom bands have $\nu = 1$, while the two middle bands, with $\nu=-1$, are connected at the Dirac points. At $t'=1$, the top two as well as the bottom two bands are gapped. As $t'$ is decreased, this gap shrinks and finally closes at $t'=0.717$. The existence and locations of Dirac points can be experimentally measured in optical lattice systems\cite{esslinger}.

\begin{figure}
\includegraphics[scale=0.85]{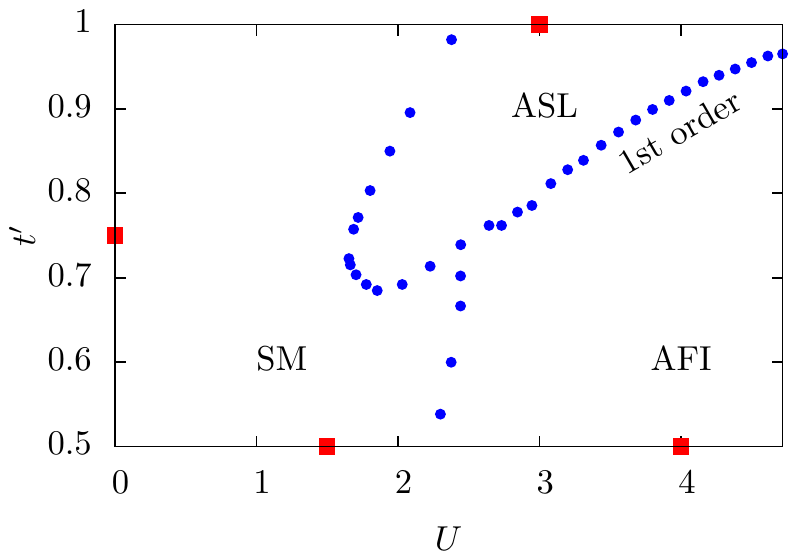}
\caption{\label{fig:phase_diagram}
(color online) The phase diagram of the Kitaev-Hubbard model at half-filling, showing the phases. 
The transition from the AFI to the ASL phase is discontinuous. The red squares correspond to the 
parameter values at which the spectral graphs have been plotted in Fig.~\ref{fig:spectral}.
}
\end{figure}

At $t'=t$ and in the large $U$ limit, the model \eqref{eq:ham} is analytically tractable. We will show that, in this regime, (a) TR symmetry is satisfied, (b) the spin-spin correlation function has power-law behavior, and (c) the corresponding spin liquid remains gapless in the spin excitations. In addition, using the variational cluster approximation (VCA) \cite{Potthoff:2003b} and cluster perturbation theory (CPT) \cite{Senechal:2000}, we have delineated a region on the $U-t'$ plane where the staggered magnetization vanishes and the spectral gap is nonzero. As this region includes the $t=t'$ line above a certain critical $U$ value, we surmise that it constitutes an ASL phase. 

CPT and VCA allow us to map the spectral gap of the model onto the $t'-U$ plane and to calculate the extent of the N\'eel phase. VCA also allows us to find out whether or not the transitions out of the N\'eel phase are continuous. However, the same cannot be done for the Mott transitions to the spin-liquid phases for which a cluster dynamical mean-field technique would be required\cite{jeanpaul}.

\begin{figure}
\includegraphics[width= \hsize]{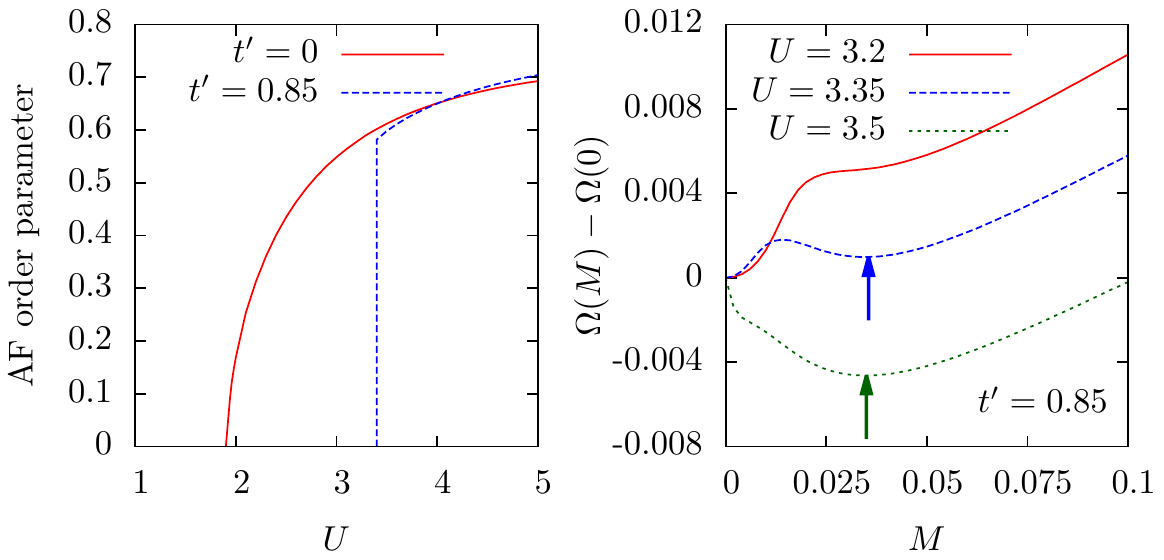}
\caption{(color online) Left panel: AF order parameter computed in VCA for $t'=0.85$ and $t'=0$ as a function of $U$. The transition is discontinuous in the first case, and continuous at $t'=0$.
Right panel: profile of the Potthoff functional as a function of Weiss field $M$ for three values of $U$ across the transition at $t'=0.85$, demonstrating the first-order character of the magnetic transition there. Arrows indicate the positions of the minima, associated with magnetic solutions, metastable in one case ($U=3.35$) and stable in another ($U=3.5$).}
\label{fig:orderp}
\end{figure}

\begin{figure}
\includegraphics[width=\hsize]{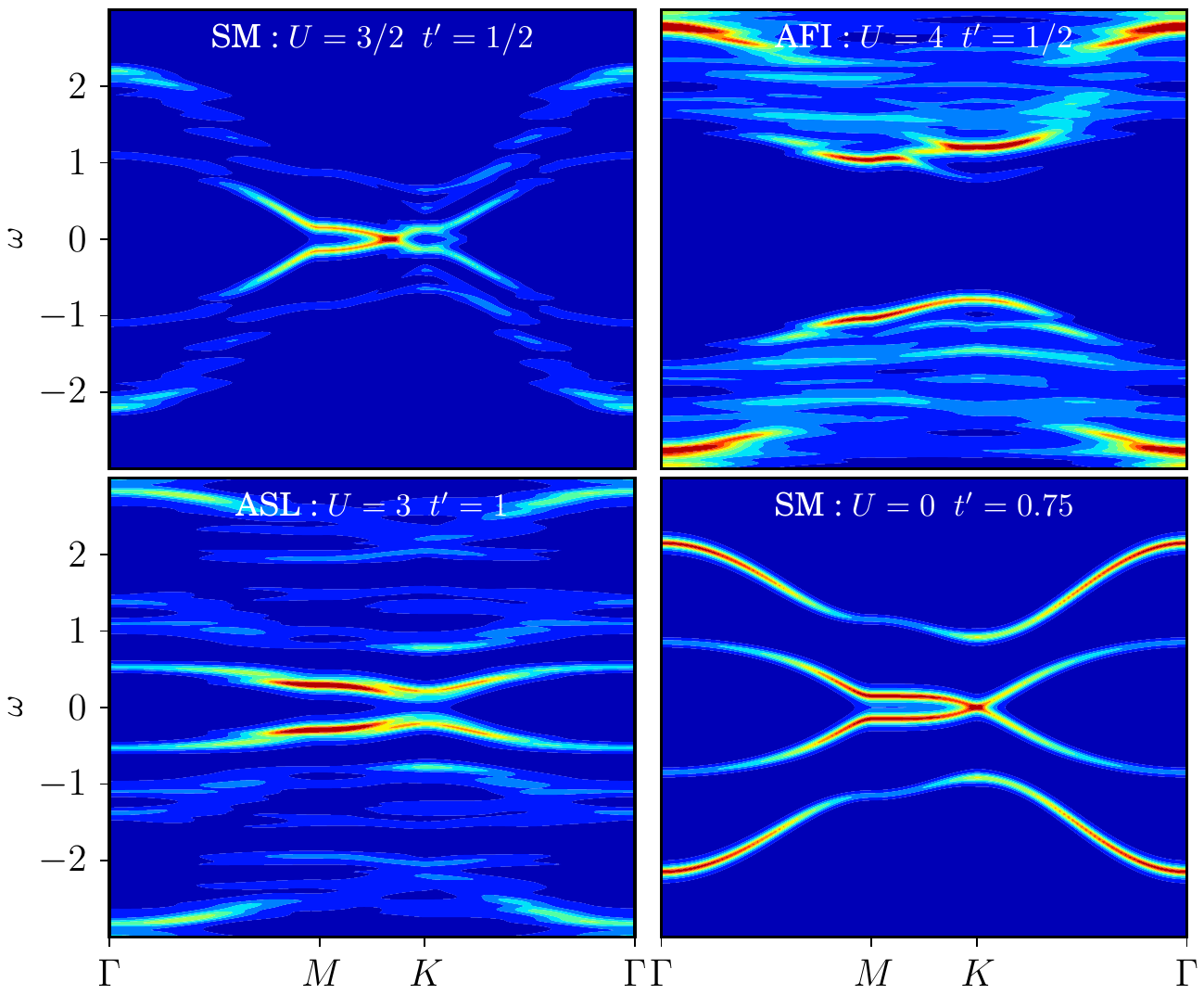}
\caption{Spectral functions of the Kitaev-Hubbard 
model, computed using CPT, as a function of energy ($\omega$, y-axis) and momentum ($k$, x-axis) 
for the four sets of parameter values marked by the squares in 
Fig.~\ref{fig:phase_diagram}, red indicating maximum value and blue indicating minimum values.
The spectrum is gapless only for the SM. 
$\Gamma$, $M$ and $K$ represent the high symmetry points of the Brillouin zone.}
\label{fig:spectral}
\end{figure}

The phase diagram of the Kitaev-Hubbard model from $t'=1$ to $t'=0.5$ is summarized in Fig.~\ref{fig:phase_diagram} (we set $t=1$). At low $U$ value, there is a TR-breaking semi-metallic phase (SM), characterized by gapless charged, spin-1/2 fermionic quasiparticles.  This nonmagnetic phase exists in the region $1 > t' > 0.5$ and $U \lesssim 2.4$. When $U \approx 2.4$ and $t'=1$, a spectral gap opens up, signature of a Mott transition from  the SM to the ASL phase, which extends to $U\to\infty$. Between $U \gtrsim 1.5$ and $U \lesssim 2.4$ and with steadily decreasing $t'$, the system starts off in the SM phase, then makes a transition into the ASL phase and finally reenters the SM phase until $t' = 0.5$. For $U>2.4$, decreasing $t'$ destabilizes the ASL phase and brings about  a transition to the antiferromagnetic (N\'{e}el) phase (AFI) which also has a spectral gap.

The ASL phase is bounded by the AFI and SM phases and is hence not connected to the possible short-ranged spin liquid at $t'=0$\cite{Meng:2010kx,sorella,jeanpaul}. The SM and AFI phases do not have quasiparticles with fractional quantum numbers or statistics. Thus, the ASL is topologically distinct from the SM and the AFI, and we expect the transitions between them to be discontinuous, as illustrated for instance in Fig.~\ref{fig:orderp}.

Spectral functions illustrating each of the three phases and computed with CPT are shown in Fig.~\ref{fig:spectral}. It is intriguing that the single-particle bands of opposite Chern numbers remain gapped in the same range of $t'$ as the existence of the ASL. This seems to indicate that geometric phase effects may play an important role in this model.

Let us now proceed to the large-$U$ analytic treatment of the model on the $t'=1$ line. We study the ASL spin-spin correlations by deriving an effective, large-$U$ spin Hamiltonian of the Kitaev-Hubbard model at half-filling. To leading order in $1/U$, this effective Hamiltonian is
\begin{eqnarray} \label{heff2} H^{(2)}&=& \sum_{\L ij\R_\alpha} \Big[
\frac{(1-t'^2)}{U}\Sv_{i}\cdot\Sv_{j}+\frac{2t'^2}{U} S_{i}^\alpha
S_{j}^\alpha\Big] \end{eqnarray}
This is a combination of the Heisenberg and Kitaev models similar to the one studied earlier\cite{Chaloupka:2010uq,schaffer2012}. As mentioned earlier, this Hamiltonian continues to have short-ranged spin-spin correlations. We have calculated the next order term in the effective spin Hamiltonian and have obtained 
\begin{align}
\label{heff4}
H^{(4)} &= \sum_{\substack{\langle ij\rangle_\alpha \\ \beta\neq\alpha}}\bigg[ \frac{(t^{'4}-1)}{U^3} {\bf{S}}_{i}.{\bf{S}}_j 
-\frac{2t^{'4}}{U^3}S_{i}^\alpha S_{j}^\alpha \nonumber \\
&-\frac{2t^{'2}}{U^3}(S_{i}^\alpha S_{j}^\beta +  S_{j}^\alpha S_{i}^\beta ) \bigg]+\sum_{\langle\langle ij\rangle\rangle_{\alpha\beta}} \bigg[ \frac{(1-t^{'2})^2}{4U^3} {\bf{S}}_{i}.{\bf{S}}_{j} \nonumber \\
&+\frac{t^{'2}-t^{'4}}{2U^3}(S_{i}^\alpha S_{j}^\alpha + S_{i}^\beta S_{j}^\beta) + 3\frac{t^{'2}}{U^3}S_{i}^\alpha S_{j}^\beta \bigg]
\end{align}
where $\L ij\R_\alpha$ denote nearest neighbors in the $\alpha$ direction and $\L\L ij\R\R_{\alpha\beta}$ denote next-nearest neighbors reached by first moving in the $\alpha$ direction and then in the $\beta$ direction.

We show below that the terms in Eq. (\ref{heff4}) induce algebraic spin-spin correlations. The effective Hamiltonian $H^{(2)}+H^{(4)}$ is comprised of two-spin operators and is hence TR symmetric. Using the charge-conjugation symmetry of the microscopic Hamiltonian (\ref{eq:ham}), we have analytically proved that the effective spin model is TR symmetric to all orders in $1/U$ (see Supplementary  Material). If this emergent symmetry in the Mott phase is not spontaneously broken then the spinons are forced to be gapless.

We now compute the spin-spin correlation,
\begin{equation}
\label{spcorrdef}
g({\bf r},t) = 
\left\langle T\left(S_{{\bf r}l}^\alpha(t)S_{\mathbf{0} m}^\beta(0)\right)\right\rangle
\end{equation}
where $\rv$ is a site of the Bravais lattice and $l$ is the sublattice index. We compute this perturbatively in the fermionic representation of the spins \cite{kitaev}, in which the Majorana fermion operators $c_i$ and $b_i^\alpha$ are defined as follows:
\begin{eqnarray}
\nonumber
\sigma_i^\alpha = ic_ib_i^\alpha,&~~~&\{c_i,c_j\}=2\delta_{ij}\\ 
\label{mfdef}
\{b^\alpha_i,b^\beta_j\}=2\delta_{\alpha\beta}\delta_{ij}, 
&~~~&\{c_i,b^\alpha_j\}=0.
\end{eqnarray}
The physical subspace is defined by the constraint
\begin{equation}
c_ib_i^xb_i^yb_i^z\vert \psi\rangle_{\rm phys}=\vert\psi\rangle_{\rm phys}
\end{equation}
In terms of these Majorana fermions, the leading order Hamiltonian is
\begin{equation}
\label{fermham}
\mathcal{H}_0=J\sum_{\langle ij\rangle_\alpha}ic_ic_jib^\alpha_ib^\alpha_j
\end{equation}
where $J=(1+t^{\prime 2})/U-(1+t^{\prime 4})/U^3$.
This Hamiltonian describes Majorana fermions ($c_i$), which we refer to 
as spinons, propagating in the 
background of static $Z_2$ gauge fields 
($u_{\langle ij\rangle_\alpha}=ib^\alpha_ib^\alpha_j$).
The spin-spin correlation functions therefore factorize into 
propagators of the $c_i$ operators. Since the spin operators create two
units of flux on adjoining plaquettes, the Majorana fermion propagators are
in the background of an even number of fluxes at a few points. 

We are interested in the asymptotic form of the leading order correction, $g^{(2)}({\bf r},t)$ (second order), when $|\rv|,~t\rightarrow\infty$. Tikhonov {\it et al.} \cite{Tikhonov:2011kx} have shown that in this limit, the propagators are the same as those in the flux-free background. Their result can be physically understood by noting that the particles hopping far way from the flux pairs will not pick up any phases from them. Thus, we can expect the long wavelength modes to be insensitive to a few localized flux pairs. 

To compute the asymptotic form or the propagators, we can derive the continuum theory of the low-energy modes in the flux-free background. The Hamiltonian (\ref{fermham}), when $ib^\alpha_ib^\alpha_j=1$, reduces to nearest-neighbor hopping on a honeycomb lattice just as in graphene. Graphene has low-energy Dirac quasiparticles about two points, ${\bf K}$ and ${\bf K'}$, in the Brillouin zone. However, since the $c$ operators are Hermitian $c_k^\dagger = c_{-k}$, the excitations only exist over half the Brillouin zone. Thus, the low-energy modes constitute a single Dirac quasiparticle. The continuum theory is derived by introducing slowly varying fields $\psi_l({\bf r})$ such that
\begin{align}
\label{psidef}
c_{{\bf r}l}(t) = \frac{1}{2}\left(
e^{i{\bf K}\cdot{\bf r}} \psi_l({\bf r},t) 
+e^{-i{\bf K}\cdot{\bf r}} \psi^\dagger_l({\bf r},t)\right)
\end{align}
$\psi({\bf r},t)$ satisfies the gapless Dirac equation. Any mass term for a {\em single} Dirac fermion breaks TR invariance. Thus the gaplessness is protected by the TR invariance of the effective spin model.

We have calculated the dynamical spin-spin correlations in the long-wavelength limit following the methods of Tikhonov \etal\cite{Tikhonov:2011kx} (see Supplemental Material) in terms of the products of the noninteracting $c$-fermion propagators $G(r,t,0,0)$
\begin{align} \label{G}
G(r,t,0,0) & =(\boldsymbol{\tau}\cdot\rv -Jt\mathbb{I} ) \frac{1}{4\pi}\frac{1}{(\rv^2 -J^2t^2)^{\frac{3}{2}} }.
\end{align}
as
\begin{align}
\label{spincorr}
\langle S^z_{\rv l}&(t)S^z_{\mathbf{0}l}(0)\rangle \nonumber\\
&= \left( -0.56\cos(2\Kv\cdot\rv)\gamma_1^2+1.13\gamma_1\gamma_2 + 0.88\gamma_2^2\right)\det G  \nonumber\\
& +0.28 \gamma_1^2 {\rm tr}(\tau_xG\tau_x G) +(0.56\gamma_1^2 +0.16\gamma_1\gamma_2){\rm tr}(G\tau_x G) \nonumber\\
& +(0.28\gamma_1^2+0.02\gamma_2^2+0.16\gamma_1\gamma_2) ({\rm tr}G)^2
\end{align}
where $\gamma_1=-2t^{\prime 2}/U^3,~\gamma_2=3t^{\prime 2}/2U^3$, and $\Kv$ is the Dirac point ($2\pi/3,2\pi/3\sqrt3$). Here, $t$ is the time, $\boldsymbol{\tau}=(\tau^x,~\tau^y)$ are the Pauli spin matrices.

Using Eqs. (\ref{spincorr}) and (\ref{G}), we find that the long-wavelength correlation function falls off as $1/r^4$. This exponent is the same as the one computed by single-spin perturbations studied in Ref.~\onlinecite{Tikhonov:2011kx} and can be motivated by simple dimensional counting. This proves the existence of the ASL in the Kitaev-Hubbard model. Although the prefactor is extremely small for large $U$ ($\sim 1/U^6$), this is the leading behavior at long distances. Therefore the effect of the perturbation cannot be neglected for any value of $U$, however large. Indeed,  we can expect the strength of these correlations to grow as $U$ decreases.

Thus, at large $U$, the leading order contribution to the spin susceptibility is independent of $U$ as in the Kitaev model, whereas the next order contribution goes as $(t/U)^6$.  The $U$ dependence of the spin susceptibility will hence be of the form $\chi=a+b(t/U)^6$, where $a$ and $b$ are constants independent of $U$. Experimental methods for measuring the spin susceptibility in cold atom systems have recently been developed~\cite{ketterle}. The value of $(t/U)^6$, for the lowest values of $(t/U)$ that the ASL exists, ranges from $0.08-0.005$, depending on $t'$. Thus, susceptibility measurements as a function of $U$, with an accuracy of about $1\%$, can provide evidence for the existence of the ASL in this model.

\begin{figure}
\includegraphics[width=0.4\textwidth]{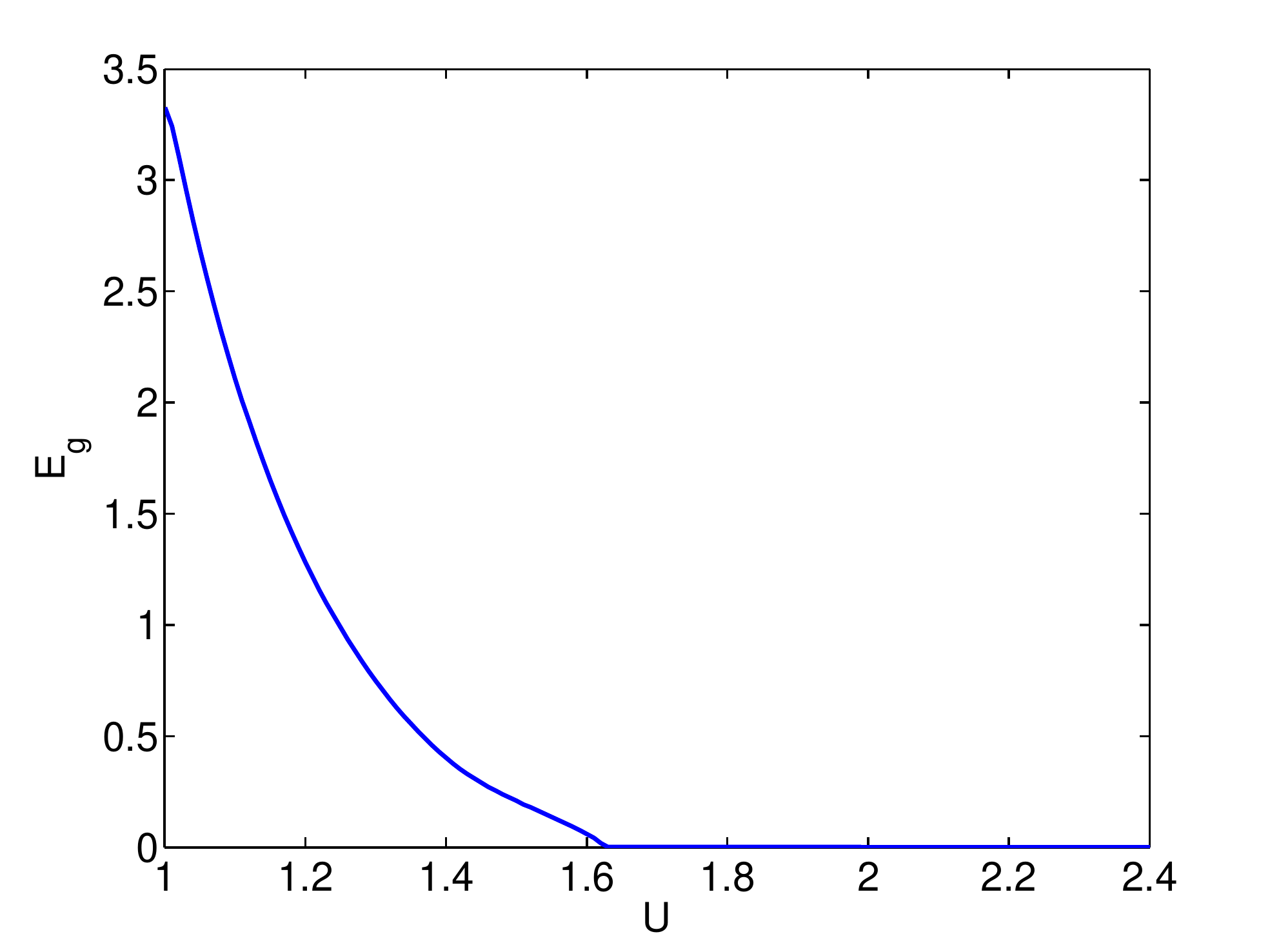}
\caption{(color online) Gap of the spinon spectrum as a function of $U$ for $t' = 1$}
\label{fig:gap}
\end{figure}

The stability of the ASL in this model comes from the preservation of TR symmetry. We have investigated the possibility of spontaneous breaking of TR symmetry and the consequent emergence of a chiral spin liquid (CSL) with a spinon gap (see Supplemental Material). In the fermionized version\cite{kitaev} of the effective spin model, where $H_{\rm eff}=H^{(2)}+H^{(4)}$, we use a mean-field theory in which the vison  and the spinon sectors are decoupled. This mean-field theory is exact for the Kitaev model, which is obtained by putting $t' = 1$ in $H^{(2)}$.  We find that the CSL solutions occur only for $U\lesssim 1.6$ and for $0.5\leq t'\leq 1$ and, thus, are not seen in the Mott regime $U\gtrsim 2.4$. In Fig.~\ref{fig:gap}, we plot the spinon gap as a function of $U$ for $t'=1$.

In conclusion, we have shown that the Kitaev-Hubbard model shows a Mott transition from a semi-metallic phase to an algebraic spin-liquid phase. The former breaks time-reversal symmetry whereas the latter preserves it. The ASL is stabilized by TR symmetry. We have proved the TR invariance in the Mott phase (to all orders in $t/U$), using charge-conjugation symmetry. At intermediate $U$, the ASL phase occurs for a wide range of $t'$, which narrows down as $U$ is increased.  Concrete schemes to realize this model have been proposed\cite{Duan:2003dq,zoller}, and experimental methods to probe the semi-metal at low $U$ values\cite{esslinger} and the ASL at large $U$ values \cite{ketterle} exist. This demonstration of the existence of the ASL might help achieve a better understanding of the physics of the pseudogap phase of the underdoped high-temperature superconductors\cite{wen1996,rantner2001,senthil2005}.


We thank G. Baskaran, Jean Paul Latyr Faye, Andr\'e-Marie Tremblay and Mukul Laad for useful discussions and R. Adhikari for feedback on the manuscript.  Computational resources were provided by Compute Canada, Calcul Qu\'ebec, and Annapurna IMSc.


\section{Supplemental Material}

\section{Cluster Perturbation theory and the VCA}

Cluster Perturbation Theory (CPT) is an approximation scheme for the one-electron Green function $\Gv(\omega)$ within Hubbard-like models.\cite{Senechal:2000, Senechal:2002,Senechal:2011vn}
It proceeds by dividing the infinite lattice $\gamma$ into a super-lattice $\Gamma$ of identical clusters of $L$ sites each (Fig.1 of the main text illustrates the cluster used in this work).
The lattice Hamiltonian $H$ is written as $H = H_c + H_T$, where $H_c$ is the cluster Hamiltonian, obtained by severing the hopping terms between different clusters, which are put into $H_T$.
Let $\Tv$ be the matrix of inter-cluster hopping terms and ${\Gv^{c}}(\omega)$ the exact Green function of the cluster.
Because of the periodicity of the super-lattice, $\Tv$ can be expressed as a function of the reduced wave-vector $\kvt$ and as a matrix in site indices within the cluster: $T_{ab}(\kvt)$. Likewise, ${\Gv^{c}}$ is a matrix in cluster site indices only, since all clusters are identical: $G^c_{ab}(\omega)$. Thus, hopping matrices and Green functions in what follows will be $\kvt$-dependent matrices of order $L$, the number of sites within each cluster.
The fundamental result of CPT for the system's one-electron Green function is 
\begin{equation}\label{eq:CPT1}
\Gv^{-1}(\kvt,\omega)={\Gv^{c}}^{-1}(\omega)-\Tv(\kvt)
\end{equation}
In practice ${\Gv^{c}}(\omega)$ is calculated numerically by the Lanczos method and the cluster must be small enough for this to be possible.
Because the lattice tiling breaks the original translation invariance of the lattice, a prescription is needed to restore the translation invariance of the resulting Green function. 
The CPT prescription for this periodization is
\begin{equation}
G(\kv,\omega) = \frac1{L}\sum_{a,b} \er^{-i(\kv)\cdot(\rv_a-\rv_b)} G_{ab}(\kv,\omega)
\end{equation}
where now $\kv$ belongs to the Brillouin zone of the original lattice and the sum is carried over cluster sites.
This formula is exact in both the strong $(t\to 0)$ and the weak ($U\to 0$) coupling limits. 

Once the approximate interacting Green function can be calculated, the spectral function $A(\kv,\omega) = -2\,{\rm Im}\, G(\kv,\omega)$ follows.
From there the density of states $N(\omega)$ can be calculated by numerically integrating $A(\kv,\omega)$ over wave-vectors
and the existence of a spectral gap can be assessed. Numerically, the density of states is always evaluated at a complex frequency with a small imaginary part $\eta$ that broadens the spectral peaks. By applying a few values of $\eta$ and extrapolating to $\eta\to0$, one can detect the presence (or not) of a spectral gap at the Fermi level. This allows us to distinguish between a metal and a Mott insulator.

If we know at what wave-vector the gap first opens up, as is the case at $t'=0$ (The Dirac points), then we can estimate the gap more reliably, without the need to extrapolate to $\eta\to0$, by simply looking up the Lehmann representation of the CPT Green function, which can be calculated when the cluster Green function is computed using the band Lanczos method. 

The Variational Cluster Approximation (VCA) is an extension of CPT in which parameters of the cluster Hamiltonian $H_c$ may be treated variationally, according to Potthoff's Self-Energy Functional Theory (SFT).\cite{Potthoff:2003b,Potthoff:2012fk}
In particular, it allows the emergence of spontaneously broken symmetries and provides an approximate value for the system's grand potential $\Omega$. 
In the case at hand, a single variational parameter is used: the strength $M_c$ of a staggered magnetization field that is added to the cluster Hamiltonian:
\begin{equation}
H_M = M\sum_\alpha  m_\alpha c^\dagger_\alpha c_\alpha 
\end{equation}
where the symbol $m_\alpha$ is $+1$ for spin-up orbitals on the A sublattice and spin-down orbitals on the B sublattice, and $-1$ otherwise.

Technically, VCA proceeds by minimizing the following quantity:
\begin{equation}
\label{omegadef}
\Omega(M_c) = \Omega_c(M_c)
- \int\frac{\dr\omega}{\pi}\frac{\dr^2k}{(2\pi)^2}\sum_\kvt \ln\det\left[\mathbf{1}-\Tv(\kvt)\Gv(\kvt,i\omega)\right]
\end{equation}
where $\Omega_c(M_c)$ is the grand potential of the cluster alone (obtained in the exact diagonalization process).
The integral over frequencies is carried over the positive imaginary axis.
At the optimal value $M_c^*$, $\Omega(M_c^*)$ is the best estimate of the system's grand potential.
At this value of $M_c$, the order parameter $M$ is calculated:
\begin{equation}
M = \int \frac{\dr^2\tilde k}{(2\pi)^2}\int\frac{\dr\omega}{2\pi}  m_\alpha G_{\alpha\alpha}(\tilde\kv,i\omega)
\end{equation}
where $G_{\alpha\alpha}$ are the diagonal elements of the CPT Green function (\ref{eq:CPT1}).

VCA provides estimates of order parameters, much like mean-field theory, but is quite superior to it because the Hamiltonian remains fully interacting (no factorization of the interaction) and spatial correlations are treated exactly within the cluster.

In this work, the VCA was used to find the phase boundary of the antiferromagnetic phase.
On the other hand, the transition between the spin liquid and semi-metal phases was found by monitoring the closure of the gap via CPT only.

\section{Proof of time reversal symmetry in the Mott Phase}

In this section we will outline the derivation of the  effective
spin Hamiltonian and prove that it is time reversal (TR) invariant in the Mott phase.

\subsection{Notation}

The Hamiltonian is taken to be:
\begin{eqnarray}
\label{kh}
H&=&\mathcal{H}_0+\mathcal{H}_K\\
\mathcal{H}_{0}&=&U\sum_{i}n_{i\uparrow}n_{i\downarrow}\\
\mathcal{H}_K&=&\sum_{\langle ij\rangle_a} 
\left\{c^\dagger_i\left(\frac{t+t'\sigma_a}{2}\right) c_j+\mathrm{H.c}\right\}
\end{eqnarray}
Let $P_m$ represent the projector on the Hilbert subspace containing $m$ 
doubly occupied sites. Thus $P_m^2=P_m$ and $\sum_{m=0}^\infty P_m=1$.

\par We write the Hamiltonian as
\begin{eqnarray}
H &=& \mathcal{H}_0+T_0+T_1+T_{-1}\\
T_s&=&\sum_{m=0}^\infty P_{s+m}\mathcal{H}_KP_m \qquad (s=-1,0,1) \\
 P_{-1}&\equiv& 0
\end{eqnarray}
$T_{-s}=T_s^\dagger$ and $T_{s}$ increases the doubly occupied sites by $s$. 
Thus we have
\begin{align}
[\mathcal{H}_{0},T_{s}] &= sUT_s\label{T0} 
\end{align}
The operators that commute with $\mathcal{H}_0$ are called {\em block diagonal} (BD) and 
the others {\em off block diagonal} (OBD).

\subsection{Canonical Perturbation Theory}

We follow the approach of Chernyshev \etal\cite{chernyshev2004} and
block diagonalize the Hamiltonian order by order in $t/U$. 
The $(k+1)^{th}$ order Hamiltonian is written as
\begin{equation}
\mathcal{H}^{(k+1)} = e^{S^{(k)}} e^{S^{(k-1)}} e^{S^{(k-2)}} \cdots H \cdots e^{-S^{(k-2)}} e^{-S^{(k-1)}}e^{-S^{(k)}}\label{ct}
\end{equation}
$S^{(k)}$ is chosen so as to eliminate the OBD terms of order $\left(\frac{t}{U}\right)^k$ 
that survive in the Hamiltonian after performing the canonical transformation at order $k-1$. 
By construction, $S^{(k)}$ does not contain terms that preserve the number of doubly occupied sites. 

$S^{(k)}$ has to satisfy the equation
\begin{align}
[S^{(k)},\mathcal{H}_0] & = -\mathcal{H}^k_{\rm OBD}
\end{align}
where $\mathcal{H}^{(k)}_{\rm OBD}$ is the OBD part of $\mathcal{H}^{(k)}$. It is easy to show
that 
\begin{align}
S^{(k)} = \sum_{i\neq j} \frac{1}{U(i-j)} P_{i} \mathcal{H}^{(k)} P_{j}\label{S}
\end{align}

The effective spin Hamiltonian at half-filling, $H^{(k)}$, is
obtained by projecting $\mathcal{H}^{(k)}$ onto the singly occupied subspace.

\subsection{Charge conjugation and Time reversal symmetries}

The Hamiltonian \eqref{kh} is symmetric under charge conjugation (C)
(particle-hole) transformation:
 \begin{align}
U_C~c_{i\sigma}^{\dagger}~U_C^\dagger& =  \eta_i c_{i\sigma}
\end{align}
where $\eta_i$ is $+1$ on sublattice A and $-1$ on sublattice B.
The unitary operator $U_C$ can be explicitly written as
\begin{align}
U_C&=\prod_ie^{i\pi S^y_i}e^{i\pi G^y_i}
\end{align}
where $S_i^a$ are the spin operators acting on the singly occupied states
and $G_i^a$ are the pseudo-spin operators acting on the empty and doubly
occupied states, and defined as

\begin{equation}
G_i^z= \frac{1}{2}\left(n_{\uparrow i}+n_{\downarrow i}-1\right)
\qquad 
G_i^+= c^\dagger_\uparrow c^\dagger_\downarrow = (G_i^-)^\dagger
\end{equation}
Every term in the Hamiltonian, $\mathcal{H}_0$ and $T_s$, is C invariant:
\begin{equation}
U_CH U_C^\dagger=H,\qquad U_C T_s U_C^\dagger=T_s
\end{equation}
It then follows from equations (\ref{S}) and (\ref{ct}) that
every term of $H^k$ is C invariant, for all $k$.

The time reversal operator is
\begin{equation}
U_T~c_{i\sigma}^{\dagger}~U_T^\dagger =  i\sigma^y_{\sigma\sigma'} c_{i\sigma'}\qquad\qquad
U_T =\prod_j e^{i\pi S^y_j}\mathcal{K}
\end{equation}
where $\mathcal{K}$ is the complex conjugation operator.
Any state $|\rm hf\rangle$ in the singly occupied subspace satisfies
the condition $G_i^a|\rm hf\rangle=0$. It then follows that
\begin{equation}
U_CU_T|\rm hf\rangle=\mathcal{K}|\rm hf\rangle
\end{equation}

\subsection{Time reversal symmetry of $H^{(k)}$}

We show the TR symmetry of $H^{(k)}$ by explicitly showing the equality
of the matrix elements of $H^{(k)}$ and  $U_T H^{(k)} U^\dagger_T$ in a real
basis. Specifically, we can choose the simultaneous eigenstates of $S^z_i$,
\begin{equation}
S^z_i\vert\{\sigma_i\}\rangle=\frac{1}{2}\sigma_i\vert\{\sigma_i\}\rangle
\end{equation}
We can always choose $S^z_i$ to be real and hence we have 
$\mathcal{K}\vert\{\sigma_i\}\rangle=\vert\{\sigma_i\}\rangle$. It
then follows that,
\begin{align}
\nonumber
\langle \{\sigma_i\}| U_T H^{(k)} U_T^\dagger  |\{\sigma_i\}'\rangle &=
\langle\{\sigma_i\}| U_T^\dagger U_C^\dagger U_C H^{(k)} U_C^\dagger U_C U_T |\{\sigma_i\}'\rangle\\
\nonumber
& = \langle\{\sigma_i\}| \mathcal{K} H^{(k)} \mathcal{K} |\{\sigma_i\}' \rangle\\
& = \langle\{\sigma_i\}|  H^{(k)} |\{\sigma_i\}' \rangle
\end{align}
Thus the effective spin Hamiltonian is TR symmetric. This implies that it
does not contain any odd-spin terms.

\section{Spin-Spin Correlation function}
We now outline the computation of the spin-spin correlation function.
We write the Hamiltonian as
\begin{align}
\label{effh}
H & = \mathcal{H}_0 + \mathcal{H}_{p} \\
\mathcal{H}_0 &= J\sum_{\langle ij\rangle_\alpha}S_{i}^{\alpha}S_{j}^{\alpha}\\
\mathcal{H}_p &= \sum_{\substack{\langle ij\rangle_\alpha\\ \alpha\neq\beta}}\delta_1 S_{i}^\beta S_j^\beta  +\gamma_1\left[ S_{i}^\alpha S_{j}^\beta +  S_{j}^\alpha S_{i}^\beta \right] \nonumber\\
&+\sum_{\langle\langle ij\rangle\rangle_{\alpha\beta}} [  \delta_2{\bf{S}}_{i}.{\bf{S}}_{j} +\delta_3(S_{i}^\alpha S_{j}^\alpha + S_{i}^\beta S_{j}^\beta)+ \gamma_2S_{i}^\alpha S_{j}^\beta ]
\end{align}
where
\begin{align}
J&=\left(\frac{1+t^{'2}}{U}-\frac{1+t^{'4}}{U^2}\right) &\gamma_1&=-\frac{2t^{'2}}{U^3}\nonumber\\
\gamma_2&=\frac{3t^{'2}}{U^3} & \delta_1 &= \frac{t^{'4}-1}{U^3} \nonumber\\
\delta_2 &= \frac{(1-t^{'2})^2}{4U^3} & \delta_3 &= \frac{t^{'2}-t^{'4}}{2U^3}
\end{align}
We take the Hamiltonian $\mathcal{H}_0$ (the Kitaev model) as the unperturbed 
Hamiltonian and $\mathcal{H}_{p}$ as a perturbation. 
We want to compute the correlation function
\begin{align}
\label{sscorr}
g({\bf r},t) = 
\left \langle T\left(S_{{\bf r}l}^\alpha(t)S_{\mathbf{0} m}^\beta(0)\right)\right\rangle
\end{align}
Where ${\bf r} = r_1 {\bf{e}}_1 + r_2 {\bf{e}}_2$, ${\bf{e}}_1$ and 
${\bf{e}}_2$ are basis vectors as shown in Fig.(\ref{fig:basis}) and $l,m$ are
the sub-lattice indices. 
To leading order, this is the spin-spin correlation function of the Kitaev 
model. The  Kitaev model has a 6-spin conserved operator associated with every 
plaquette, $W_p$ which can take values $\pm 1$ and can be 
interpreted as a $Z_2$ flux.\cite{kitaev} The ground state is in the flux-free sector 
($W_p=1~\forall p$). The spin operators at site $\rv$ create a pair of flux tubes 
in two of the plaquettes that the site belongs to. Since the time evolution
does not change the flux configuration, the spin-spin correlation 
(\ref{sscorr}) is zero except when $\rv$ and $\mathbf{0}$ are nearest neighbors.\cite{Mandal:2011ys} 

The second-order perturbation term is
\begin{align}
\label{g2}
g^{(2)} & = \frac{(-i)^2}{2}\int d\tau_1\int d\tau_2 \nonumber\\
&\left\langle T\left( S_{{\bf{r}} l}^\alpha(t)\mathcal{H}_p(\tau_1)\mathcal{H}_p(\tau_2)
S_{0 m}^\beta(0)\right)\right\rangle.
\end{align}
The time evolution is governed by $\mathcal{H}_0$. This term will be non-zero
only if there are terms in $\mathcal{H}_p$ such that the product of the four
operators in (\ref{g2}) do not change the flux configuration of the ground
state.\cite{Mandal:2011ys} We find that such terms do exist in $\mathcal{H}_p$.  
We concentrate on correlation function 
$\langle S_{r_1,r_2,A}^z S_{0,0,A}^z\rangle$. The following terms combine
with $S_{r_1,r_2,A}^z$ to produce flux-free configurations when acting
on the ground state,
\begin{align}
&\gamma_2 S_{r_1-1,r_2,B}^x S_{r_1-1,r_2+1,B}^y ;&\gamma_2 S_{r_1+1,r_2-1,A}^y S_{r_1+1,r_2,A}^x ;\nonumber\\
&\gamma_1S_{r_1,r_2,A}^x S_{r_1-1,r_2+1,B}^y; &\gamma_1S_{r_1,r_2,A}^yS_{r_1-1,r_2,B}^x ;\nonumber\\
&\gamma_1S_{r_1,r_2,B}^y S_{r_1+1,r_2,A}^x ;&\gamma_1S_{r_1,r_2,B}^x S_{r_1+1,r_2-1,A}^y \nonumber
\end{align}
These and the terms with $(r_1,r_2)\rightarrow (0,0)$ which combine with
$S^z_{0,0,A}$ give 36 possibly non-zero contributions to $g^{(2)}$. 

The problem now is to compute the resulting 6-spin correlation functions in 
the Kitaev model. We do this in the fermionic representation of the 
spins in an enlarged Hilbert space,\cite{kitaev} in which the Majorana fermion operators
$c_i$ and $b_i^\alpha$ are defined as follows:
\begin{eqnarray}
\label{mfdef}
\sigma_i^\alpha = ic_ib_i^\alpha,~~~~~~~\{c_i,c_j\}=2\delta_{ij} \nonumber\\
\{b^\alpha_i,b^\beta_j\}=2\delta_{\alpha\beta}\delta_{ij}, 
~~~~\{c_i,b^\alpha_j\}=0
\end{eqnarray}
The physical subspace is defined by the constraint
\begin{equation}
c_ib_i^xb_i^yb_i^z\vert \psi\rangle_{\rm phys}=\vert\psi\rangle_{\rm phys}
\end{equation}
In terms of these Majorana fermions, the leading order Hamiltonian is,
\begin{equation}
\label{fermham}
\mathcal{H}_0=J\sum_{\langle ij\rangle_\alpha}ic_ic_jib^\alpha_ib^\alpha_j
\end{equation}
This Hamiltonian describes Majorana fermions ($c_i$), which we refer to 
as spinons, propagating in the 
background of static $Z_2$ gauge fields ($u_{\langle ij\rangle_\alpha}=ib^\alpha_ib^\alpha_j$)
The correlation function in equation (\ref{g2}) thus factorizes into 
propagators of the $c_i$ operators. Since the spin operators create two
units of flux on adjoining plaquettes, the Majorana fermion propagators are
in the background of an even number of fluxes at a few points. 

We are interested in the asymptotic form of $g^{(2)}({\bf r},t)$ when $|\rv|,~
t\rightarrow\infty$. Tikhonov {\it et. al.} \cite{Tikhonov:2011kx} have shown that in this limit, the propagators are the same as those in the flux-free 
background. Their result can be physically understood by noting that the 
particles hopping far way from the flux pairs will not pick up any phases
from them. Thus we can expect the long wavelength modes to be insensitive
to a few localized flux pairs. 

To compute the asymptotic form or the propagators, we can derive the 
continuum theory of the low-energy modes in the flux-free background.
The Hamiltonian (\ref{fermham}), when $ib^\alpha_ib^\alpha_j=1$, reduces
to nearest-neighbor hopping on a honeycomb lattice just as in graphene. 
Graphene has low-energy Dirac quasi-particles about two points, ${\bf K}$
and ${\bf K'}$, in 
the Brillouin zone. However, since the $c$ fermions are Majorana fermions,
the excitations exist only over half the Brillouin zone. Thus the low-energy
modes constitute a single Dirac quasi-particle. The continuum theory is
derived by introducing slowly varying fields $\psi_l({\bf r})$ such that
\begin{align}
\label{psidef}
c_{{\bf r}l} = \frac{1}{2}\left(
e^{i{\bf K}\cdot{\bf r}} \psi_l({\bf r}) 
+e^{-i{\bf K}\cdot{\bf r}} \psi^\dagger_l({\bf r})\right)
\end{align}
Substituting
equation (\ref{psidef}) in equation (\ref{fermham}) it can be seen
that the low energy continuum theory is that of a single Dirac fermion.
The propagator is defined as
\begin{align}
G_{lm}({\bf r},t) = 
\langle T\left(\psi_l({\bf r},t) \psi^\dagger_m(0,0)\right)\rangle
\end{align}
It can be computed to be
\begin{align}
G_{lm} & = ( {\boldsymbol{\tau}}\cdot{\bf{r}}-Jt \mathbb{I})_{lm} \frac{1}{4\pi} \frac{1}{(\rv^2 -J^2t^2)^{\frac{3}{2}} }
\end{align}
where $\boldsymbol{\tau} = (\tau_x,\tau_y)$ are the Pauli matrices.
We can then compute the correlation function in equation (\ref{g2})
to obtain the following expression:
\begin{widetext}
\begin{align}
\langle S_{{\bf{r}} l}^z(t)S_{0 l}^z(0)\rangle& =(-0.56\cos(2{\Kv}\cdot{\rv})\gamma_1^2+1.13\gamma_1\gamma_2 +1.69\gamma_2^2\epsilon)\det G \nonumber\\
& +(0.28\gamma_1^2+0.07\gamma_2^2+0.84\gamma_1\gamma_2\epsilon+0.63\gamma_2^2\epsilon^2-0.28\gamma_1\gamma_2 -0.42\gamma_2^2\epsilon) ({\rm Tr}G)^2\nonumber\\
& +0.28\gamma_1^2{\rm Tr}(\tau_xG\tau_xG)+(0.56\gamma_1^2 -0.28\gamma_1\gamma_2+0.84\gamma_1\gamma_2\epsilon){\rm Tr}(G\tau_x G) \label{spincorr}
\end{align}
\end{widetext}
where $l$ represents the sublattice index (A or B), $\alpha$ represents the 
three types of bonds $x,y,z$ and $\epsilon$ is the energy density of the 
Kitaev model. We can obtain the $\langle S_{{\bf{r}} l}^x(t)S_{0 l}^x(0)\rangle$ and $\langle S_{{\bf{r}} l}^y(t)S_{0 l}^y(0)\rangle$ from the above correlation function by using the following property: a $2\pi/3$ rotation about a sublattice $A$ point takes $x$ link to $y$ link, $y$ link to $z$ and $z$ link to $x$, in a cycle. The direction is reversed for sublattice $B$.  
Thus we can see that the \eqref{spincorr} goes as $r^{-4}$ which shows that 
we have an algebraic spin liquid (ASL) for large $U$.

\section{Mean Field Theory of the effective spin model}
The effective spin Hamiltonian is given in equation (\ref{effh}).
To investigate the instability of the ASL to CSL, we perform a mean-field treatment of the above Hamiltonian in the fermionic representation (\ref{mfdef}). 
The decoupling of the spinon and gauge field sectors is represented by
\begin{align}
\sigma_i^\alpha\sigma_j^\beta=-ic_ic_j~ib^\alpha_ib^\beta_j
&\approx -ic_ic_jB^{\alpha\beta}_{ij}-iC_{ij}b_i^\alpha b_j^\beta+C_{ij}B^{\alpha\beta}_{ij}
\end{align}
The self-consistency equations are
\begin{align}
B^{\alpha\beta}_{ij}\equiv\langle ib^\alpha_ib^\beta_j\rangle \qquad\qquad
C_{ij}\equiv\langle ic_ic_j\rangle
\end{align}

\begin{figure}[ht]
\includegraphics[scale=1]{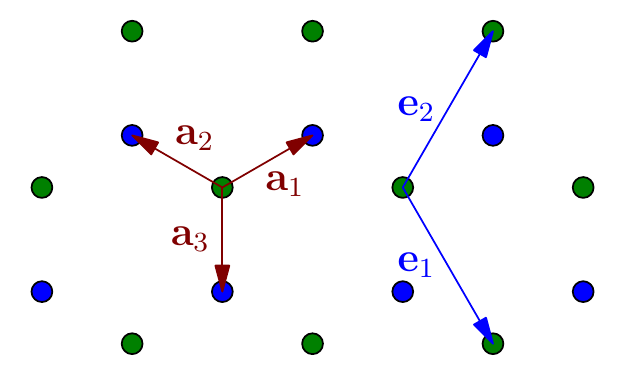}
\caption{Basis vector used. Green (Blue) represents sublattice A (B).}
\label{fig:basis}
\end{figure}

We assume that the ground state is translationally invariant, isotropic and denote
\begin{align}
\label{ccxyz}
C_{i,i+\vec{a}_\alpha}=\epsilon 
\qquad B_{i,i+\vec{a}_\alpha}^{\alpha \alpha}=\eta  
\qquad C_{i,i\pm \vec{e}_i,l}=\mu_l  \\
B_{i,i+\vec{a}_\alpha}^{\alpha \beta}=B_{\alpha}^{\alpha \beta}
\qquad B_{i,i\pm \vec{e}_i,l}^{\alpha \beta}= b_l
\qquad\alpha\neq\beta
\end{align}
where $\vec{a}_\alpha$ represents the nearest neighbor vector on the $\alpha^{\textrm{th}}$ link. $\alpha,\beta$ represent the links ($x,y,z$), $l$ indicates 
the sublattice index (A or B) and $\vec{e}_i$ represents the basis vectors of the underlying Bravais lattice (see Fig: \ref{fig:basis}).

\begin{figure}[htb]
\includegraphics[width=0.5\textwidth]{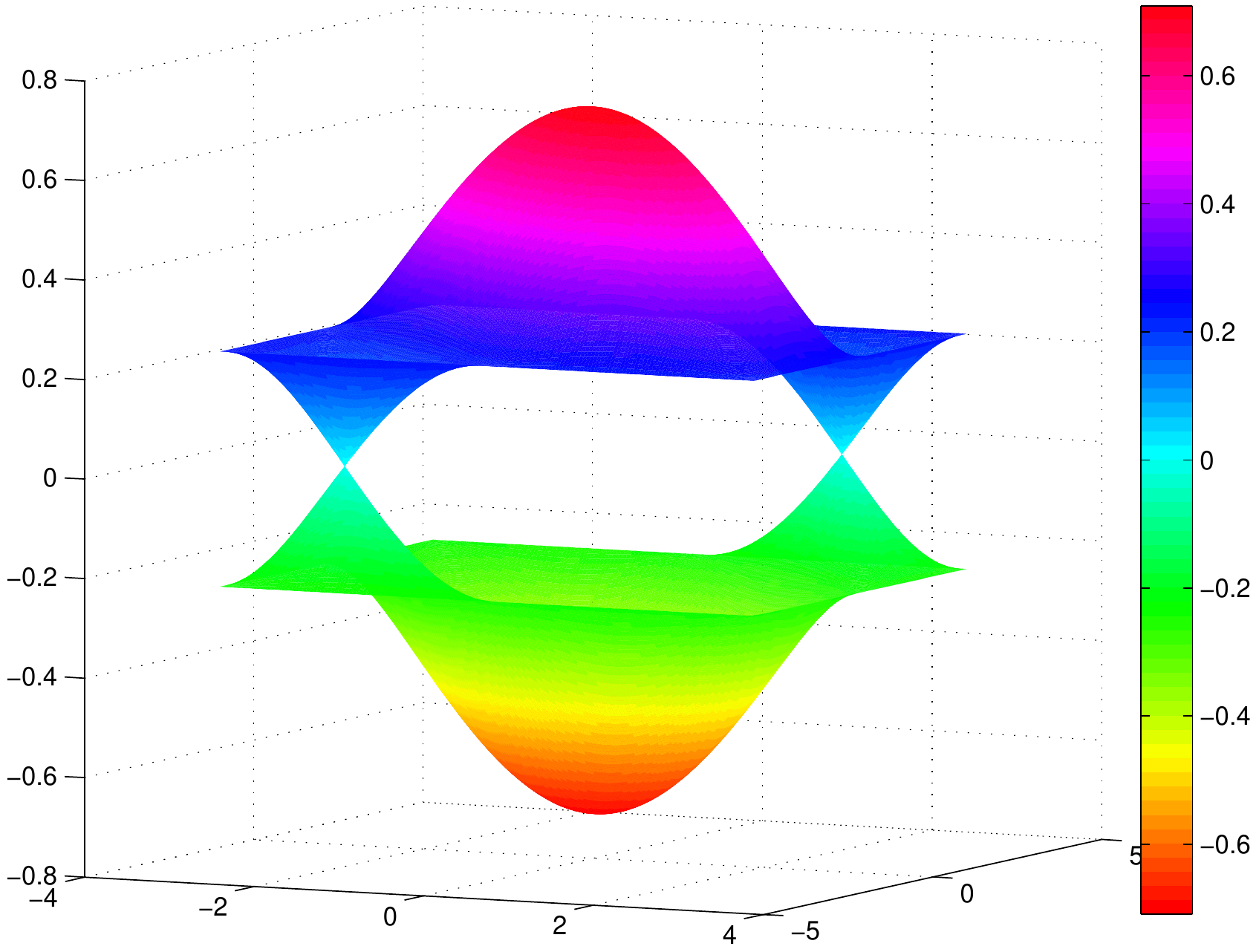}
\caption{Spinon dispersion relation at $U=2$.}
\label{c-sector}
\end{figure}

The mean field Hamiltonian at $t'=1$ is,
\begin{widetext}
\begin{align}
H_{MF}&=H_{MF}^b+H_{MF}^c\\
H_{MF}^b&=\frac{1}{4}\sum_{k\in HBZ}
\left(\begin{array}{cc}
(b^\alpha_{\kv 1})^\dagger&(b^\alpha_{\kv 2})^\dagger
\end{array}\right)     
\left(\begin{array}{cc}
iV_{\alpha\beta,1}(\kv )&iU_{\alpha\beta}(\kv )\\
-iU_{\alpha\beta}^*(\kv )&iV_{\alpha\beta,2}(\kv )   
\end{array}\right)
\left(\begin{array}{c}
b^\beta_{\kv 1}\\b^\beta_{\kv 2}
\end{array}\right)\\
U(\kv )&=\epsilon\left(\begin{array}{ccc}
Je^{-ik_1}&\gamma_1(e^{-ik_1}+e^{ik_2})&\gamma_1(e^{-ik_1} + 1)\\
\gamma_1(e^{-ik_1}+e^{ik_2})&Je^{ik_2}&\gamma_1(e^{ik_2}+1)\\
\gamma_1(e^{-ik_1}+1)&\gamma_1(e^{ik_2}+1)&J\end{array}\right)
\end{align}
\begin{align}
V_{\alpha\beta,1}=\mu_1\gamma_2\left(\begin{array}{ccc}
0&e^{ik_3}&-e^{-ik_1}\\
-e^{-ik_3}&0&e^{ik_2}\\
e^{ik_1}&-e^{-ik_2}&0
\end{array}\right)\quad
V_{\alpha\beta,2}=\mu_2\gamma_2\left(\begin{array}{ccc}
0&-e^{-ik_3}&-e^{-ik_1}\\
e^{ik_3}&0&-e^{-ik_2}\\
e^{ik_1}&e^{ik_2}&0
\end{array}\right)\\
\end{align}
\begin{align}
H_{MF}^c&=\frac{1}{4}\sum_{\kv \in HBZ}
\left(\begin{array}{cc}
c_{\kv 1}^\dagger&c_{\kv 2}^\dagger
\end{array}\right)     
\left(\begin{array}{cc}
iv_{1}(\kv )&iu(\kv )\\
-iu^*(\kv )&iv_{2}(\kv )   
\end{array}\right)\left(\begin{array}{c}
c_{\kv 1}\\
c_{\kv 2}
\end{array}\right)\\
u(\kv)&=\sum_\alpha e^{-i\vec{k}\cdot\vec{e}_\alpha} \left(J\eta +\gamma_1\sum_{\beta \neq \alpha}B_{\alpha}^{\alpha \beta} \right) 
\end{align}
\begin{align}
v_{1}(\kv)=2ib_1\gamma_2 \sum_\alpha \sin(\vec{k}\cdot\vec{e}_\alpha) 
\qquad v_{2}(\kv)&=-2ib_2\gamma_2\sum_\alpha \sin(\vec{k}\cdot\vec{e}_\alpha)
\end{align}
\end{widetext}
where $k_i = \vec{k}\cdot\vec{e}_i$

The nearest-neighbor term in Hamiltonian \eqref{effh} results in closing the
gap in the spinon sector at the Dirac points (see Fig: \ref{c-sector}), whereas the next-nearest term
collapses the gap at $(0,0)$ and $(\pi,\pi)$. For large values of $U$ the
nearest-neighbor term dominates; for smaller values of $U$, on the other hand,
the next-nearest neighbor term comes into play and the Dirac points shift to
$(0,0)$ and $(\pi,\pi)$. Numerically we find that the spinon sector is gap-less
for $U\geq 1.6$ (Fig.~5 of paper). We have checked that this remains true for $1<t'<0.5$. VCA
indicates a Mott transition at $U=2.4$. This shows the absence of the CSL phase
in the presence of higher order perturbative terms and indicates the ASL phase
continues till the Mott transition.

\end{document}